\documentclass[reprint,amsmath,amssymb,aps,floatfix,twocolumn,prb,superscriptaddress,longbibliography,nofootinbib]{revtex4-2}
\usepackage{subfigure}
 \usepackage[colorlinks=true]{hyperref}
 \usepackage{amsmath,amssymb,mathtools}
\usepackage{xcolor}
\usepackage{graphicx}
\usepackage{dcolumn}
\usepackage{bm}
\allowdisplaybreaks
\usepackage{orcidlink}
\usepackage[normalem]{ulem}

\begin{document}

\title{Gross-Neveu-Yukawa SO(2) and SO(3) tensorial criticality}

\author{SangEun Han\,\orcidlink{0000-0003-3141-1964}
}
\affiliation{Department of Physics, Simon Fraser University, Burnaby, British Columbia, Canada V5A 1S6}
\author{Shouryya Ray\,\orcidlink{0000-0003-4754-0955}
}
\affiliation{Department of Science and Technology, University of the Faroe Islands, Vestara Bryggja 15,
FO-100 T\'{o}rshavn, Faroe Islands}
\affiliation{CP3-Origins, University  of  Southern  Denmark, Campusvej  55, 5230 Odense M, Denmark}
\author{Igor F.~Herbut\,\orcidlink{0000-0001-5496-8330}
}
\affiliation{Department of Physics, Simon Fraser University, Burnaby, British Columbia, Canada V5A 1S6}
\affiliation{Institute for Solid State Physics, University of Tokyo, Kashiwa 277-8581, Japan}

\date{\today}
\onecolumngrid

\begin{abstract}
We investigate the relativistic SO(2)- and SO(3)-invariant Gross-Neveu-Yukawa field theories for real, rank-two, symmetric, traceless tensor order parameters coupled to $N_{\text{f}}$ flavors of two-component Dirac fermions. These field theories arise as an effective description of fractionalized spin-orbital liquids. The two theories are the simplest and special cases of the more general class of field theories with SO($N$) symmetric tensor order parameter coupled to Dirac fermions, in which the symmetry is low enough to allow only one, and not the usual two quartic self-interaction terms. Using a two-loop renormalization group near the upper critical dimension, we demonstrate that the theory exhibits a new critical fixed point for $N=3$ and the concomitant continuous phase transition for any value of $N_{\text{f}}$. For $N=2$ the theory is equivalent to the chiral XY model. We discuss the crucial role of the symmetry-allowed sextic self-interactions in the selection of the ground state configuration in the case of SO(3). The universal quantities such as the the anomalous dimensions of order parameters and fermions, the correlation length exponent, and the mass gap ratio between order parameter and fermion masses are computed up to $\epsilon^{2}$ order.
\end{abstract}

\maketitle

\section{Introduction}
Gross-Neveu field theory in $D = 3$ spacetime dimensions has proven to be a fertile subject of study, both in condensed-matter and statistical physics, as well as quantum field theory \cite{Gross1974,Zinn-Justin2021,Rosenstein1989,Vasilev1993,GRACEY1994a,GRACEY1994b,Erramilli2023,ZINNJUSTIN1991}. When generalized and applied to condensed matter, it describes quantum phase transitions in a plethora of semimetallic systems \cite{Herbut2006,Herbut2009a,Herbut2009b,Bitan2013,*Bitan2016e,Zerf2017,Ihrig2018,VOJTA2000,Huh2008,Schwab2022,Scherer2016,Torres2018}. The presence of gapless fermionic degrees of freedom makes these transitions distinct from their classical counterparts, while the fact that these degrees of freedom are restricted to isolated points in momentum space makes the effective field theory analytically tractable. In fact, the Gross-Neveu family of universality classes constitutes arguably one of the few well-understood universality classes beyond Ginzburg-Landau to date \cite{Herbut2024book,*Igor2024}. This in turn has made it a fruitful testing ground for quantum field theory, particularly in its bosonized (aka Yukawa) formulation, where it provides one of the few tractable examples of UV finiteness with both bosonic and fermionic field content (without supersymmetry) in the form of asymptotic safety \cite{braun11}.

Tensorial universality classes---i.e., universality classes where the order parameter transforms in a rank-2 irreducible representation of the spontaneously broken symmetry---are fascinating in their own right \cite{Herbut2023,Han2024a}. Unlike the conventional vector (= rank-1) cousin, the purely bosonic tensorial $\operatorname{SO}(N)$ relativistic field theory for $N \geqslant 4$ does not possess, in the language of the renormalization group (RG), a critical fixed point \cite{Delfino2015,Pelissetto2018}. This well-known problem arises because, unlike in the vector version, in this class of theories there are two independent quartic interactions, called ``trace'' and ``double-trace'' in field-theory literature. Both are canonically marginal at the upper critical dimension $D = 4$ and become relevant with engineering dimension unity at the physical number of spacetime dimensions $D=3$. The coupling of their RG flows leads only to runaway behavior, and presumably to fluctuation-induced first-order transitions \cite{Herbut2007}.

The runaway RG flows in the bosonic $\operatorname{SO}(N)$ tensorial field theories persist even when the order parameter is Yukawa-coupled to a limited number of Dirac fermions \cite{Herbut2022,Liu2022,*Liu2022E,Uetrecht2023}.  However, as pointed out by two of us \cite{han2024b}, upon coupling to sufficiently many flavors of Dirac fermions transforming in the fundamental representation of $\operatorname{SO}(N)$, the stable critical fixed fixed point unavoidably emerges at any $N$. This may be understood intuitively: in the presence of many fermion flavours, the order parameter is more appropriately thought of as a composite field bilinear in the fermions, which in turn behave as the true fundamental degrees of freedom. Canonically, a bosonic quartic interaction is then an octic fermionic interaction, which has canonical dimension $(-2)$ in $D=3$ spacetime dimensions. This fact renders these interactions highly irrelevant and serves to stabilize the critical fixed point.

The runaway flow of the trace and double-trace quartic couplings does not arise for the $\operatorname{SO}(N)$ tensorial field theories when $N=2,3$: in this case there is in fact only one independent quartic interaction and the critical fixed point is stable, like in the vectorial case. However, a different subtlety arises: the critical field theory can be shown to be identical to the $\operatorname{SO}(N_\text{s})$-vector theory with $N_\text{s} = \frac12 (N-1)(N+2)$ as the number of order-parameter components. This observation can be used to show order-by-order in an expansion in $\epsilon$ around the upper critical dimension, $D=4-\epsilon$, that the leading critical exponents are identical in $\operatorname{SO}(N_\text{s})$-vector and $\operatorname{SO}(N)$-tensor universality classes. This non-trivial equivalence between the vector and tensor critical points provides one motivation for the present work: coupling the tensor order parameter to fermions, as we here demonstrate, removes the equivalence, so that the tensorial $\operatorname{SO}(N)$ critical fixed point defines a genuinely different universality from its vector $\operatorname{SO}(N_\text{s})$ cousin for $N=3$. We also find that, unlike in the general case with $N\geq 4$, this new universality class exists for any number of fermion flavors, and then determine its universal characteristics, such as critical exponents, at two-loop order.

This paper also has a phenomenological motivation that goes beyond intrinsic issues of the theory of quantum criticality: critical exponents of the Gross-Neveu-SO(3) type with one fundamental Majorana fermion (or half a fundamental Dirac fermion) were proposed \cite{Seifertetal20} to describe the quantum phase transition from an SO(3) spin-orbital liquid to a conventionally ordered antiferromagnet. The Majorana fermions, called spinons, arise from fractionalization of the spin-orbital moments and present a defining characteristic of the exotic spin-orbital liquid phase, but are not directly visible to standard spectroscopic techniques. The onset of antiferromagnetism is readily detected by conventional means, and the critical exponents serve as ``fingerprints'' of the emergent Majorana fermions. It was pointed out by one of us \cite{ray2024} that the critical exponents of the tensor version of Gross-Neveu-SO(3) theory hence presents an additional, independent set of fingerprints of the same Majorana fermions. Providing theoretical predictions for these exponents, as we do here beyond the leading quantum corrections, is hence also relevant to the study of fractionalized phases of matter, especially in frustrated magnets.

This paper is organized as follows. In Sec.~\ref{sec:model} we define the GNY theory of the SO($N$) symmetric traceless tensor coupled to two-component Dirac fermions, and discuss the special cases of $N=2,3$. In Sec.~\ref{sec:two-loop} we discuss the RG flow to two loops, and display the critical points. How the RG-irrelevant sextic terms in the theory select the ground state is discussed in Sec.~\ref{sec:sextic}, and the critical exponents are computed in Sec.~\ref{sec:universal_quantities}. Finally, we present our conclusions in Sec.~\ref{sec:conclusion}.

\section{Model}\label{sec:model}

In this section, we review the Lagrangian for $\operatorname{SO}(N)$-tensor Gross-Neveu-Yukawa theory for $N = 2,3$. %
The field content comprises $N_\text{s} = \frac12 (N-1)(N+2)$ real boson components $\varphi_i$ and $2 N N_\text{f}$ complex fermion components $\psi_\alpha$. The bosonic fields are collected into a real traceless symmetric matrix $S=\sum_{i=1}^{N_\text{s}}\varphi_{i}\mathbb{S}_{i}$, where the $\mathbb{S}_i$ themselves are symmetric and traceless and obey the orthonormal property $\operatorname{Tr}[\mathbb{S}_{i}\mathbb{S}_{j}]=\bar{N}_{\text{Tr}}\delta_{ij}$, with arbitrary $\bar{N}_{\text{Tr}}$. Under rotation by $O \in \operatorname{SO}(N)$, the field $S$ transforms as $S \mapsto O S O^\intercal$. %
The fermion components are arranged as $\psi = (\psi_\alpha) = (\psi_{aI})$. Here, $a = 1,\ldots,N$ is a fundamental index, i.e., for the same $O \in \operatorname{SO}(N)$, $\psi_{aI} \mapsto O_{ab} \psi_{bI}$. The index $I = 1,\ldots,2N_\text{f}/d_\gamma$ counts the number of fermions; $N_\text{f}$ will be referred to as the number of flavors. Finally, $d_\gamma$ refers to the dimension of the Clifford algebra. For instance, in the fundamental spinor representation in integer spacetime dimension $D$, $d_\gamma = 2^{\lfloor D/2 \rfloor}$. For the analytic continuation to arbitrary $D$, $d_\gamma$ is left unspecified in intermediate steps, the Clifford algebra itself formally being infinite-dimensional; for a well-defined theory, $d_\gamma$ needs to cancel in final expressions. 
The action for the GNY field  theory is then given by
\begin{widetext}
\begin{align}
\mathcal{S} &= \int d^D x\, \left[\bar{\psi} (\mathbb{I}_{2NN_\text{f}/d_\gamma}\otimes \gamma^\mu)\partial_\mu \psi + \bar{g}\bar{\psi} (\mathbb{I}_{2 N_\text{f}} \otimes S) \psi + \frac{1}{2\bar{N}_{\text{Tr}}} \operatorname{Tr}[(\partial_\mu S)^2 + \bar{r}S^2]  + \frac{\bar{\lambda}_{1}}{4\bar{N}_{\text{Tr}}^{2}}(\operatorname{Tr} [S^2])^2+ \frac{\bar{\lambda}_{2}}{4\bar{N}_{\text{Tr}}} \operatorname{Tr} [S^4]\right].
\label{eq:model}
\end{align}
\end{widetext}
where $\mathbb{I}_{n}$ is the $n\times n$ identity matrix and repeated indices are summed over, and `$\otimes$' is a tensor product between matrices. The $\gamma^\mu$ satisfy the Clifford algebra $\{ \gamma^\mu , \gamma^\nu \} = 2 \delta^{\mu\nu} \mathbb{I}_{d_\gamma}$and $\mu = 0,\ldots,D-1$ is the spacetime index (i.e., such that $\delta^{\mu}_{\mu} = D$). Thus, $\mathbb{I}_{2N_{\text{f}}/d_{\gamma}}\otimes\gamma^{\mu}$ and $\mathbb{I}_{2N_{f}}\otimes S$ are $2NN_{\text{f}}\times2NN_{\text{f}}$ matrices, respectively.

The conjugate of the fermion field is defined as $\bar{\psi}=\psi^{\dagger}\gamma^{0}$. 
The coupling $\bar{g}$ is the Yukawa coupling and the quartic couplings $\bar{\lambda}_{1,2}$ are called trace and double-trace respectively. The ``overbar'' for couplings refers to the fact that these are dimensionful, $\bar{g} \sim (\text{length})^{(D-4)/2}$, and $\bar{\lambda}_{1,2} \sim (\text{length})^{D-4}$. It was shown in \cite{han2024b} that there is no critical fixed point in the theory for $N > 3$ unless $N_\text{f}$ is large enough. Critical fixed point is defined as being unstable only in the direction of the tuning parameter for the transition $\bar{r}$. In the present case, however, the trace and double-trace interactions are \emph{not} independent, but obey the relation
\begin{align}
\frac{1}{\bar{N}_{\text{Tr}}^{2}}\left(\operatorname{Tr} [S^{2}]\right)^{2}={}&\frac{2}{\bar{N}_{\text{Tr}}^{2}}\operatorname{Tr} [S^{4}] =\left(\sum_{i=1}^{N_{\text{s}}}\varphi_{i}^{2}\right)^{2}, \label{eq:identity24}
\end{align}
so that the two quartic interactions are reducible to a single interaction,
\begin{align}
\frac{\bar{\lambda}_{1}}{\bar{N}_{\text{Tr}}^{2}}(\operatorname{Tr} [S^{2}])^{2}+\frac{\bar{\lambda}_{2}}{\bar{N}_{\text{Tr}}}\text{Tr} [S^{4}]
={}&\frac{\bar{\lambda}}{\bar{N}_{\text{Tr}}^{2}}(\operatorname{Tr} [S^{2}])^{2},\label{eq:single_quartic}
\end{align}
where
\begin{align}
\bar{\lambda}={}&\bar{\lambda}_{1}+\frac{\bar{N}_{\text{Tr}}}{2}\bar{\lambda}_{2}.
\end{align}
Hence, the bosonic sector of the GNY theory is at this level indistinguishable from the $\operatorname{SO}(N_\text{s})$ vector theory: it is the Yukawa coupling to the fermions which encodes that the $\varphi_{i}$ ($i=1,\cdots,N_\text{s}$) in fact transform as components of a rank-2 tensor.
Note that the action for $N=2$ is equivalent to the chiral XY model \cite{Bitan2013,*Bitan2016e,Zerf2017}.

\begin{figure*}
\centering
    \subfigure[]{\includegraphics[width=0.11\linewidth]{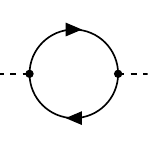}}\hfill
    \subfigure[]{\includegraphics[width=0.11\linewidth]{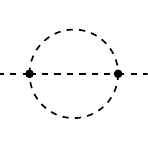}}\hfill
    \subfigure[]{\includegraphics[width=0.11\linewidth]{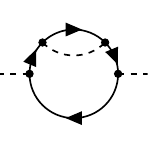}}\hfill
    \subfigure[]{\includegraphics[width=0.11\linewidth]{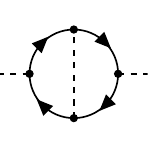}}\hfill
    \subfigure[]{\includegraphics[width=0.11\linewidth]{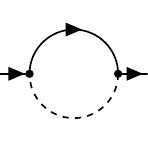}}\hfill
    \subfigure[]{\includegraphics[width=0.11\linewidth]{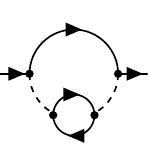}}\hfill
    \subfigure[]{\includegraphics[width=0.11\linewidth]{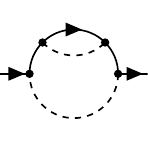}}\hfill
    \subfigure[]{\includegraphics[width=0.11\linewidth]{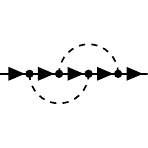}}\\
    \subfigure[]{\includegraphics[width=0.11\linewidth]{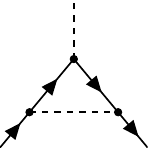}}\hfill
    \subfigure[]{\includegraphics[width=0.11\linewidth]{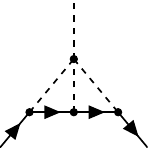}}\hfill
    \subfigure[]{\includegraphics[width=0.11\linewidth]{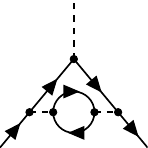}}\hfill
    \subfigure[]{\includegraphics[width=0.11\linewidth]{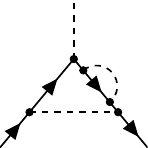}}\hfill
    \subfigure[]{\includegraphics[width=0.11\linewidth]{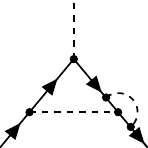}}\hfill
    \subfigure[]{\includegraphics[width=0.11\linewidth]{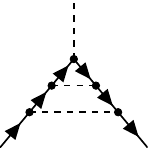}}\hfill
    \subfigure[]{\includegraphics[width=0.11\linewidth]{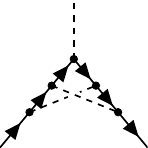}}\hfill
    \subfigure[]{\includegraphics[width=0.11\linewidth]{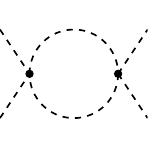}}\\
    \subfigure[]{\includegraphics[width=0.11\linewidth]{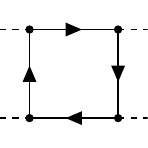}}\hfill
    \subfigure[]{\includegraphics[width=0.11\linewidth]{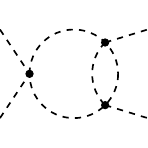}}\hfill
    \subfigure[]{\includegraphics[width=0.11\linewidth]{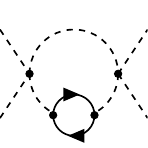}}\hfill
    \subfigure[]{\includegraphics[width=0.11\linewidth]{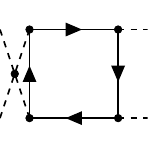}}\hfill
    \subfigure[]{\includegraphics[width=0.11\linewidth]{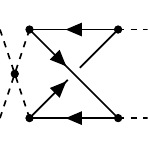}}\hfill
    \subfigure[]{\includegraphics[width=0.11\linewidth]{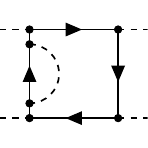}}\hfill
    \subfigure[]{\includegraphics[width=0.11\linewidth]{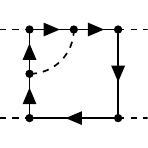}}\hfill
    \subfigure[]{\includegraphics[width=0.11\linewidth]{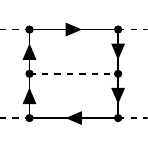}}
    \caption{Feynman diagrams illustrating loop corrections up to two-loop order for order parameter self-energy (a–d), fermion self-energy (e–h), Yukawa interaction (i–o), and quartic interactions (p–x). The solid lines with arrowheads represent fermions, while the dashed lines denote order parameters. The ordering of diagrams are following Ref.~\cite{Jack2024}.
    }\label{fig:feynman}
\end{figure*}

\section{RG flow at two-loop}\label{sec:two-loop}
The action defined in Eq.~\eqref{eq:model} in $D = 4-\epsilon$ spacetime dimensions falls in the class of perturbatively renormalizable boson-fermion theories whose beta functions and anomalous dimensions for arbitrary field content have been derived in the modified minimal subtraction ($\overline{\text{MS}}$) scheme by Jack, Osborn and Steudtner \cite{Jack2024} to three-loop order. The main difficulty in applying the general results to a specific field content lies in  accounting for the fact that most couplings in the generic theory considered in \cite{Jack2024} are in fact not independent but related to each other by (possibly intricate) symmetry relations, and therefore in projecting them on to physically meaningful couplings. This has been automatized to two-loop order in the package \texttt{RGBeta} \cite{Thomsen2021}.  We  have also evaluated the (universal) two-loop beta-functions ourselves, and  found them to agree fully with the result we extracted using \texttt{RGBeta}.
To elaborate, we get the beta functions by calculating the loop correlations for the Feynman diagrams in Fig.~\ref{fig:feynman} up to two-loop order following \cite{Jack2024}. The two-loop beta-functions read
\begin{widetext}
\begin{align}
\frac{d\alpha_{g}}{d\ell}={}&\epsilon\alpha_{g}-\alpha_{g}^{2}\frac{(N^{2}+(2N_{\text{f}}+3)N-6)}{8}-\frac{(N^{2}+N+2)}{64}\alpha_{g}\lambda^{2}+\frac{(2N^{2}+3N-6)}{16}\alpha_{g}^{2}\lambda \notag\\
&
+\frac{(N^{4}+2N^{3}(12N_{\text{f}}+1)+3N^{2}(16N_{\text{f}}-9)-12N(8N_{\text{f}}+3)+36)}{512}\alpha_{g}^{3},\label{eq:rg1}\\
\frac{d\lambda}{d\ell}={}&\epsilon\lambda-\alpha_{g}\lambda\frac{NN_{\text{f}}}{2}+\frac{N^{2}N_{\text{f}}}{4}\alpha_{g}^{2}-\frac{(N^{2}+N+14)}{8}\lambda^{2}
-\frac{N^{2}N_{\text{f}}(N^{2}+6N-8)}{32}\alpha_{g}^{3} \notag\\
&+\frac{NN_{\text{f}}(3N^{2}-7N-2)}{64}\alpha_{g}^{2}\lambda+\frac{NN_{\text{f}}(N^{2}+N+14)}{32}\alpha\lambda^{2}+\frac{3(3N^{2}+3N+22)}{32}\lambda^{3}.\label{eq:rg2}
\end{align}
\end{widetext}
Here, $\ell \coloneqq \ln(\Lambda/\mu)$ is the RG time, with $\Lambda$ a reference UV scale and $\mu$ the $\overline{\text{MS}}$ renormalization scale. The dimensionless couplings appearing above are related to the dimensionful ones appearing in the action \eqref{eq:model} as $\alpha_g \coloneqq \mu^{\epsilon} \bar{g}^{2} (\bar{N}_{\text{Tr}}/N) / (2\pi^{2})$ and $\lambda \coloneqq \mu^\epsilon \bar{\lambda}/(2\pi^2)$. The scalar mass $\bar{r} = \mu^2 r$ is symmetry-allowed, but actually vanishes identically on the critical surface, $r_* = 0$, in massless renormalization schemes. The required powers of $\mu$ for a given coupling follows from its engineering dimension; the dimensionless rescaling factors are chosen for convenience. 
Note that the RG flow equations for $N=2$ are fully consistent with the chiral XY model \cite{Bitan2013,*Bitan2016e,Zerf2017}.

To solve for the fixed point $d(\alpha_g,\lambda)/d\ell|_{(\alpha_g,\lambda) = (\alpha_g^*,\lambda^*)} = 0$, we assume an asymptotic expansion $(\alpha_g^*,\lambda^*) = \sum_{n=1}^\infty (\alpha_{g,n}^*,\lambda_n^*) \epsilon^n$. To find the leading corrections, insert the ansatz into the flow equations above and compare coefficients at $O(\epsilon^2)$, which yields a decoupled quadratic equation purely in $\alpha_{g,1}^*$. The solution $\alpha_{g,1}^* = 0$ leads to a sector where the fermions decouple and the theory becomes the classical $\operatorname{O}(N_\text{s})$ Heisenberg universality class.\footnote{This fixed point is also unstable in the sense that its stability matrix has a positive eigenvalue $\epsilon$; its eigenvector points along the Yukawa coupling axis. This perturbation of course does not exist in a purely bosonic theory, such that the Wilson-Fisher fixed point is stable, and only becomes unstable once embedded within the Gross-Neveu-Yukawa theory space.} The interacting Yukawa solution $\alpha_{g,1}^* \neq 0$ inserted back into $\lambda$'s flow equation leads to a quadratic equation for $\lambda_{1}^*$. Both solutions are interacting, but only $\lambda_1^* > 0$ is stable. We therefore choose this branch. The equation for $(\alpha_{g,n}^*, \lambda_n^*)$ is found by inserting the $(\alpha_{g,k}^*,\lambda_k^*)_{k < n}$ into the fixed-point equation and comparing coefficients at $O(\epsilon^{n+1})$. The equation is always \emph{linear} in $(\alpha_{g,n}^*,\lambda_n^*)$ (with the equation for $\alpha_{g,n}^*$ in fact independent of $\lambda_{n}^*$). Hence, the higher-order corrections are always \emph{unique}, once the solution branch is selected at $O(\epsilon)$.

The critical fixed point for $N=2$ found this way is given by
\begin{widetext}
\begin{align}
\alpha_{g}^{*}={}&\frac{2}{(N_{\text{f}}+1)}\epsilon
+\frac{((2N_{\text{f}}+38)F_{2}(N_{\text{f}})-(2N_{\text{f}}^{2}-224N_{\text{f}}+137))}{100(N_{\text{f}}+1)^{3}}\epsilon^{2}+\mathcal{O}(\epsilon^{3}),\label{eq:fp2a}\\
\lambda^{*}={}&
\frac{(F_{2}(N_{\text{f}})-N_{\text{f}}+1)}{5(N_{\text{f}}+1)}\epsilon
+\frac{\epsilon^{2}}{500(N_{\text{f}}+1)^{3}F_{2}(N_{\text{f}})}\left[(36N_{\text{f}}^{4}+342N_{\text{f}}^{3}+573N_{\text{f}}^{2}-4911N_{\text{f}}+60)\right.\notag\\&\left.-(36N_{\text{f}}^{3}-342N_{\text{f}}^{2}-1449N_{\text{f}}-60)F_{2}(N_{\text{f}})\right]+\mathcal{O}(\epsilon^{3}),
\end{align}
where $F_{2}(N_\text{f})=\sqrt{N_{\text{f}}^{2}+38N_{\text{f}}+1}$.
The corresponding critical fixed point for $N=3$ is given by
\begin{align}
\alpha_{g}^{*}={}&\frac{3}{4(N_{\text{f}}+2)}\epsilon
+\frac{[(14N_{\text{f}}+336)F_{3}(N_{\text{f}})-(14N_{\text{f}}^{2}-3046N_{\text{f}}+173)]}{1014(N_{\text{f}}+2)^{3}}\epsilon^{2}+\mathcal{O}(\epsilon^{3}),\\
\lambda^{*}={}&\frac{2(F_{3}(N_{\text{f}})-N_{\text{f}}+2)}{13(N_{\text{f}}+2)}\epsilon
+\frac{\epsilon^{2}}{13182(N_{\text{f}}+2)^{3}F_{3}(N_{\text{f}})}\left[(900N_{\text{f}}^{4}+16562N_{\text{f}}^{3}+100613N_{\text{f}}^{2}-185344N_{\text{f}}+16704)\right.\notag\\&\left.-(900N_{\text{f}}^{3}-5038N_{\text{f}}^{2}-57143N_{\text{f}}-8352)F_{3}(N_{\text{f}})\right]+\mathcal{O}(\epsilon^{3}),\label{eq:fp3lambda}
\end{align}
\end{widetext}
where $F_{3}(N_{\text{f}})=\sqrt{N_{\text{f}}^{2}+48N_{\text{f}}+4}$. Note that the one-loop results in Eqs.~\eqref{eq:fp2a}-\eqref{eq:fp3lambda} agree with Ref.~\cite{Jack2024}.

\begin{figure}
    \centering
\subfigure[]{
\includegraphics[width=0.465\linewidth]{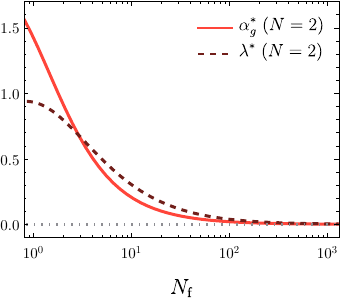}
}\hfill
\subfigure[]{
\includegraphics[width=0.465\linewidth]{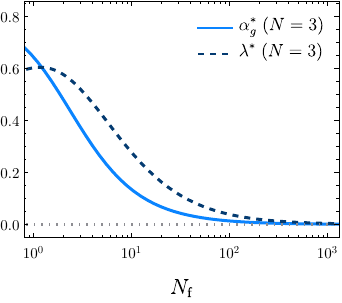}
}
    \caption{Stable interacting fixed point values of $\alpha_{g}$ (solid line) and $\lambda$ (dashed line) for (a) $N=2$ and (b) $N=3$ in terms of $N_{\text{f}}$. The fixed point values are positive for all $N_{f}$. They are positive, implying a real and bounded-from-below fixed-point action for all $N_\text{f}$. Here we set $\epsilon=1$.}
    \label{fig:fixed_point}
\end{figure}

The fixed-point values for $N=2,3$ are presented in Fig.~\ref{fig:fixed_point}. It shows that the critical fixed-point couplings $(\alpha_g^*,\lambda^*)$ remain positive at two-loop order and continue to yield a bosonic fixed-point potential that is real and bounded from below. It is also worth noticing that the two-loop terms are all of order $1/N_f ^2$ at large $N_f$.

Note that the stability matrices of the fixed points have negative eigenvalues for small $\epsilon$ but also can have positive eigenvalues for larger values of $\epsilon$ depending on $N_{f}$. This behavior arises as an artifact of the $\epsilon$-expansion and can be corrected through appropriate resummation methods, but also requires comparison with numerics \cite{Liu2024,Reehorst2023} and actual experiments. We leave this for future work.

\section{Sextic terms and the ground state}\label{sec:sextic}

The order-parameter sextic self-interactions appear irrelevant at the non-interacting fixed point for small $\epsilon$, but they would be the leading irrelevant interactions that reduce the $\text{SO}(N_s)$ symmetry of the bosonic effective action, present up to quartic terms, down to $\text{SO}(N)$. As such they select the ground state configuration of the order parameter.

For a general $N$, the possible $\text{SO}(N)$-symmetric sextic interactions of tensor order parameters are as follows:
\begin{align}
\mathcal{R}_{222}={}&\frac{1}{\bar{N}_{\text{Tr}}^{3}}(\text{Tr}[S^{2}])^{3},\\
\mathcal{R}_{24}={}&\frac{1}{\bar{N}_{\text{Tr}}^{2}}\text{Tr}[S^{2}]\text{Tr}[S^{4}],\\
\mathcal{R}_{33}={}&\frac{1}{\bar{N}_{\text{Tr}}^{2}}(\text{Tr}[S^{3}])^{2}\\
\mathcal{R}_{6}={}&\frac{1}{\bar{N}_{\text{Tr}}}\text{Tr}[S^{6}].
\end{align}
For $N=2$ and 3, the above interactions are not independent of each other.
For example, from the relation in Eq.~\eqref{eq:identity24}, since $\operatorname{Tr}[S^{4}]\propto(\varphi_{i}^{2})^{2}\propto(\text{Tr}[S^{2}])^{2}$, we have
\begin{align}
\mathcal{R}_{24}=\frac{\bar{N}_{\text{Tr}}}{2}\mathcal{R}_{222}.
\end{align}
Also, when $N=2$, $\mathcal{R}_{33} = 0$. Similarly, $\mathcal{R}_{6}$ can be written in terms of $\mathcal{R}_{222}$ and $\mathcal{R}_{33}$ as follows:
\begin{align}
\mathcal{R}_{6}={}&
\begin{cases}
\displaystyle \frac{\bar{N}_{\text{Tr}}^{2}}{4}\mathcal{R}_{222},&N=2,\\
\displaystyle \frac{\bar{N}_{\text{Tr}}^{2}}{4}\mathcal{R}_{222}+\frac{\bar{N}_{\text{Tr}}}{3}\mathcal{R}_{33}, &N=3.\\
\end{cases}
\end{align}
As a result, the irreducible sextic terms in the order-parameter potential for $N=2$ and $N=3$ can be written as
\begin{align}
V_{6}(S)=
\begin{cases}
\displaystyle \frac{\bar{\kappa}}{6}\mathcal{R}_{222},&N=2\\
\displaystyle \frac{\bar{\kappa}_{1}}{6}\mathcal{R}_{222}+\frac{\bar{\kappa}_{2}}{6}\mathcal{R}_{33},&N=3.\\
\end{cases}
\end{align}
Since $\text{Tr}[S^{2}]\propto \varphi_{i}^{2} $, $\mathcal{R}_{222}$ is still invariant under the whole SO($N_{s}$) symmetry, but $\mathcal{R}_{33}$ is invariant only under the smaller SO($N$). In a case of $N=2$, the matrix form of the possible ordered ground state $\bar{S}$ is unique, $\bar{S}\propto \text{diag}(1,-1)$, so the quartic interaction is enough to determine the ordered ground state. Let us therefore focus on $N=3$.
Depending on the sign of $\bar{\kappa}_{2}$, the symmetry of the ground state configuration will differ \cite{Boettcher2018}. If the sign of $\bar{\kappa}_{2}$ is positive, $\mathcal{R}_{33}$ should be minimized, and the resulting ground state configuration will be biaxial nematic. The SO($N$) symmetry would in this case be broken down to a discrete symmetry. However, if the sign of $\bar{\kappa}_{2}$ is negative, $\mathcal{R}_{33}$ should be maximized by the ground state configuration, which will therefore have a form of the uniaxial nematic. Consequently, the original SO(3) symmetry would be broken down to the continuous SO(2).

The computed RG flow equations, now including the sextic interactions, are given at one-loop order by
\begin{align}
\frac{d\alpha_{g}}{d\ell}={}&\epsilon\alpha_{g}-\alpha_{g}^{2}\frac{3(N_{\text{f}}+2)}{4},\\
\frac{d\lambda}{d\ell}={}&\epsilon\lambda-\frac{3N_{\text{f}}}{2}\alpha_{g}\lambda+\frac{9N_{\text{f}}}{4}\alpha_{g}^{2}-\frac{13}{4}\lambda^{2}+9\kappa_{1}+\frac{3}{2}\kappa_{2},\\
\frac{d\kappa_{1}}{d\ell}={}&-2(1-\epsilon)\kappa_{1}-\frac{9N_{\text{f}}}{4}\alpha_{g}\kappa_{1}-\frac{81N_{\text{f}}}{32}\alpha_{g}^{3}+\frac{93}{64}\lambda^{3}\notag\\&-\frac{57}{8}\lambda\kappa_{1}-\frac{9}{16}\lambda\kappa_{2},\\
\frac{d\kappa_{2}}{d\ell}={}&-2(1-\epsilon)\kappa_{2}-\frac{9N_{\text{f}}}{4}\alpha_{g}\kappa_{2}-\frac{9N_{\text{f}}}{8}\alpha_{g}^{3}-\frac{15}{4}\lambda\kappa_{2},
\end{align}
where $\kappa_{1}=\mu^{-2(1-\epsilon)}\bar{\kappa}_{1}/(2\pi^{2})$ and $\kappa_{2}=\mu^{-2(1-\epsilon)}\bar{\kappa}_{2}(\bar{N}_{\text{Tr}}/N)/(2\pi^{2})$.
We can find a critical fixed point as follows to leading order in $\epsilon$:
\begin{align}
\alpha_{g}^{*}={}&\frac{4}{3(N_{\text{f}}+2)}\epsilon,\\
\lambda^{*}={}&\frac{2(F_{2}(N_{\text{f}})-N_{\text{f}}+2)}{13(N_{\text{f}}+2)}\epsilon,\\
\kappa_{1}^{*}={}&-\frac{3\epsilon^{3}}{8788(N_{\text{f}}+2)^{3}}\left[(31 N_{\text{f}}^{3}+1023 N_{\text{f}}^{2}+6742 N_{\text{f}}-248)\right.\notag\\&\left.\quad-(31 N_{\text{f}}^{2}+279 N_{\text{f}}+124)F_{3}(N_{\text{f}})\right],\\
\kappa_{2}^{*}={}&-\frac{4N_{\text{f}}}{3(N_{\text{f}}+2)^{3}}\epsilon^{3}.
\end{align}
Due to the terms $\sim \alpha_g ^3$ the couplings $\kappa_1$ and $\kappa_2$ are generated by the integration over fermions, and both are negative and of order
$\epsilon^{3}$ at the fixed point. The negative value of $\kappa_1$ necessitates  the inclusion of higher order terms to recover the stable theory; these will be generated as well by the fermions, but will be of higher order in $\epsilon$. The leading term that breaks $SO(N_s)$ symmetry is therefore $\sim \kappa_2$.
Since the fixed-point value of $\kappa_{2}$ is also negative, $\mathcal{R}_{33}$ is maximized by the ground state configuration, which is therefore the uniaxial nematic:
\begin{align}
\bar{S}={}&\bar{\varphi}_{0}\left(\begin{matrix}
1&&\\
&1&\\
&&-2
\end{matrix}\right).
\end{align}
Note that the negative $\kappa_{2}$ also can be obtained by the mean-field analysis with the sextic interaction from the fermion loop. %

With the ground state fixed in the symmetry-broken phase, it is worth asking what this means for the fermionic spectrum. From Eq.~\eqref{eq:model}, we can read off the fermion mass matrix as $M_\psi = \sqrt{\alpha_g^*} \bar{S}$ (in units of the UV cutoff), where it is safe to set $\alpha_g = \alpha_g^*$ because the Yukawa coupling is an irrelevant coupling at the critical fixed point. The squared eigenvalues of $M$ work out to $m_{\psi,a}^2 \propto \bar{\phi}_0^2$ with two-fold degeneracy and $m_{\psi,b}^2 = 4 m_{\psi,a}^2$ with no degeneracy. Note that the spectrum is completely gapped, as opposed to the vector Gross-Neveu-Yukawa-SO(3) theory \cite{Seifertetal20}, and in agreement with mean-field theory for the Gross-Neveu-SO(3) tensor theory \cite{ray2024}.

\section{Universal quantities}\label{sec:universal_quantities}

\subsection{Critical exponents}
We find the critical exponents, such as the anomalous dimensions of fermion and boson, $\eta_{\psi}$ and $\eta_{\phi}$, and inverse correlation length exponent to be as follows:
\begin{align}
\eta_{\psi}={}&\frac{(N^{2}+N-2)}{16}\alpha_{g}\notag\\&-\frac{(N^{2}+N-2)(N^{2}+(12N_{\text{f}}+1)N-2)}{1024}\alpha_{g}^{2},\\
\eta_{\phi}={}&\frac{NN_{\text{f}}}{4}\alpha_{g}-\frac{(3N^{2}+5N-10)NN_{\text{f}}}{128}\alpha_{g}^{2}\notag\\&
+\frac{(N^{2}+N+2)}{64}\lambda^{2},\\
\nu^{-1}={}&
2-\frac{NN_{\text{f}}}{4}\alpha_{g}-\frac{(N^{2}+N+2)}{8}\lambda\notag\\&
+\frac{(3N^{2}+N-2)NN_{\text{f}}}{128}\alpha_{g}^{2}
+\frac{NN_{\text{f}}(N^{2}+N+2)}{32}\alpha_{g}\lambda\notag\\&
+\frac{5(N^{2}+N+2)}{64}\lambda^{2}.
\end{align}

When $N=2$, right at the critical  fixed point, they are given by
\begin{widetext}
\begin{align}
\eta_{\psi}
={}&\frac{1}{2(N_{\text{f}}+1)}\epsilon-\frac{[(152N_{\text{f}}^{2}-49N_{\text{f}}+162)-2(N_{\text{f}}+19)F_{2}(N_{\text{f}})]}{400(N_{\text{f}}+1)^{3}}\epsilon^{2}+\mathcal{O}(\epsilon^{3}),\\
\eta_{\phi}
={}&\frac{N_{\text{f}}}{(N_{\text{f}}+1)}\epsilon+\frac{[(112N_{\text{f}}^{2}-249N_{\text{f}}+2)+2(19N_{\text{f}}+1)F_{2}(N_{\text{f}})]}{200(N_{\text{f}}+1)^{3}}\epsilon^{2}+\mathcal{O}(\epsilon^{3}),\\
\nu^{-1}
={}&2-\frac{(4N_{\text{f}}+1+F_{2}(N_{\text{f}}))}{5(N_{\text{f}}+1)}\epsilon \notag\\
&+\frac{[(68N_{\text{f}}^{4}+4646N_{\text{f}}^{3}-576N_{\text{f}}^{2}+11732N_{\text{f}}-70)-(68N_{\text{f}}^{3}+104N_{\text{f}}^{2}+313N_{\text{f}}+70)F_{2}(N_{\text{f}})]}{1000(N_{\text{f}}+1)^{3}F_{2}(N_{\text{f}})}\epsilon^{2}+\mathcal{O}(\epsilon^{3}).
\end{align}
\end{widetext}

For example, when $N_{\text{f}}=1$,
\begin{align}
\eta_{\psi}(N=2,N_{\text{f}}=1)={}&0.25\epsilon-0.00376\epsilon^{2},\\
\eta_{\phi}(N=2,N_{\text{f}}=1)={}&0.5\epsilon+0.0737\epsilon^{2},\\
\nu^{-1}(N=2,N_{\text{f}}=1)={}&2-1.132\epsilon+0.243\epsilon^{2}.
\end{align}
The series seems well-behaved to this order, in that the coefficients are decreasing with each order.  Substituting $\epsilon=1$ we find
\begin{align}
\eta_{\psi}(N=2,N_{\text{f}}=1,\epsilon=1)={}&0.246,\\
\eta_{\phi}(N=2,N_{\text{f}}=1,\epsilon=1)={}&0.574,\\
\nu^{-1}(N=2,N_{\text{f}}=1,\epsilon=1)={}&1.110.
\end{align}
Furthermore, the critical exponents for $N_{\text{f}}=1/2,2$ are consistent with the literature \cite{Zerf2017}.

\begin{figure*}
\centering
\subfigure[]{
\includegraphics[width=0.32\linewidth]{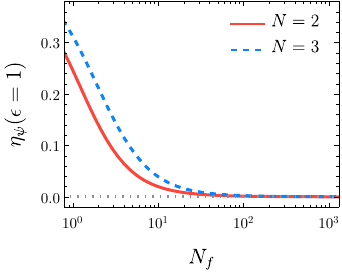}
}\hfill
\subfigure[]{
\includegraphics[width=0.32\linewidth]{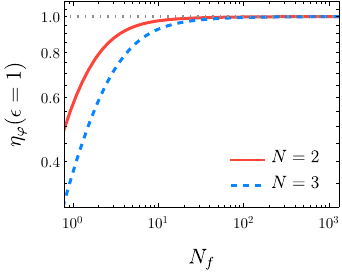}
}\hfill
\subfigure[]{
\includegraphics[width=0.32\linewidth]{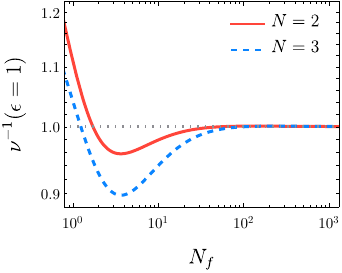}
}
\caption{Critical exponents for $N=2$ and $N=3$ up to $\epsilon^{2}$ order. We put $\epsilon=1$. (a) The anomalous dimension of fermions, (b) the anomalous dimension of order parameters, and (c) inverse correlation length exponents. The solid red line and dashed blue line stand for $N=2$ and $N=3$, respectively. The anomalous dimensions of fermions and order parameters approach 1 and 0 in the large-$N_{\text{f}}$ limit, respectively, and the inverse correlation length exponent approaches 1 in the large-$N_{\text{f}}$ limit.}
\label{fig:critical_exp}
\end{figure*}

For $N=3$, the critical exponents are given by
\begin{widetext}
\begin{align}
\eta_{\psi}
={}&\frac{5}{6(N_{\text{f}}+2)}\epsilon-\frac{5[(3084N_{\text{f}}^{2}-2209N_{\text{f}}+2209)-42(N_{\text{f}}+24)F_{3}(N_{\text{f}})]}{24336(N_{\text{f}}+2)^{3}}\epsilon^{2}+\mathcal{O}(\epsilon^{3}),\\
\eta_{\phi}
={}&\frac{N_{\text{f}}}{(N_{\text{f}}+2)}\epsilon+\frac{[(4738N_{\text{f}}^{2}-9319N_{\text{f}}+336)+168(6N_{\text{f}}+1)F_{3}(N_{\text{f}})]}{4056(N_{\text{f}}+2)^{3}}\epsilon^{2}+\mathcal{O}(\epsilon^{3}),\\
\nu^{-1}
={}&2-\frac{(7F_{3}(N_{\text{f}})+19N_{\text{f}}+14)}{26(N_{\text{f}}+2)}\epsilon
+\frac{\epsilon^{2}}{52728(N_{\text{f}}+2)^{3}F_{3}(N_{\text{f}})}\left[7(660 N_{\text{f}}^{4}+60502 N_{\text{f}}^{3}+12019 N_{\text{f}}^{2}+268960 N_{\text{f}}-10464)\right.\notag\\
&\left.-6(770 N_{\text{f}}^{3}+4505 N_{\text{f}}^{2}+13733 N_{\text{f}}+6104)F_{3}(N_{\text{f}})\right]+\mathcal{O}(\epsilon^{3})
\end{align}
\end{widetext}

For example, when $N_{\text{f}}=1$,
\begin{align}
\eta_{\psi}(N=3,N_{\text{f}}=1)={}&0.278\epsilon+0.0347\epsilon^{2},\\
\eta_{\phi}(N=3,N_{\text{f}}=1)={}&0.333\epsilon+0.0394\epsilon^{2},\\
\nu^{-1}(N=3,N_{\text{f}}=1)={}&2-1.076\epsilon+0.118\epsilon^{2}.
\end{align}
The two-loop terms for $\epsilon=1$ are again significantly smaller than the one-loop terms, and the series is well-behaved to this order. For $\epsilon=1$,
\begin{align}
\eta_{\psi}(N=3,N_{\text{f}}=1,\epsilon=1)={}&0.312,\\
\eta_{\phi}(N=3,N_{\text{f}}=1,\epsilon=1)={}&0.373,\\
\nu^{-1}(N=3,N_{\text{f}}=1,\epsilon=1)={}&1.042,
\end{align}

The critical exponents for arbitrary $N_{\text{f}}$ with $\epsilon=1$ for $N=2$ and 3 are presented in Fig.~\ref{fig:critical_exp}.

\subsection{Mass-gap ratio}

We can also consider the mass gap ratio between boson and fermion masses \cite{ZINNJUSTIN1991,han2018,han2024b}.
First of all, let us consider the ground states of the symmetry broken phase.
The possible configurations are as follows:
 \begin{align}
 S={}&\bar{\varphi}\left(\begin{matrix}
 1&0\\
 0&-1\\
 \end{matrix}\right),\;\;\bar{\varphi}^{2}=-\frac{\bar{N}_{\text{Tr}}r}{2\lambda}\text{ for $N=2$},
 \intertext{and}
S={}&\bar{\varphi}\left(\begin{matrix}
1&0&0\\
0&1&0\\
0&0&-2\\
 \end{matrix}\right),\;\;\bar{\varphi}^{2}=-\frac{\bar{N}_{\text{Tr}}r}{6\lambda}\quad\text{ for $N=3$}.
 \end{align}
The squared mass of the bosonic radial mode is given by\footnote{There are also $N - 1$ Goldstone modes.}
\begin{align}
m_{\varphi}^{2}={}&
\begin{cases}
-\frac{4r}{\bar{N_{\text{Tr}}}},&N=2\\
-\frac{12r}{\bar{N_{\text{Tr}}}}&N=3.
\end{cases}
\end{align}
The fermion mass square for $N=2$ is given by $m_{\psi}^{2}=\bar{g}^{2}\bar{\varphi}^{2}=-\bar{N}_{\text{Tr}}\bar{g}^{2}r/(2\bar{\lambda})$.
In a case of $N=3$, one finds two possible values of fermion mass:
\begin{align}
m_{\psi,a}^{2}={}&\bar{g}^{2}\bar{\varphi}^{2}=-\frac{\bar{N}_{\text{Tr}}\bar{g}^{2}r}{6\bar{\lambda}},\\
m_{\psi,b}^{2}={}&4\bar{g}^{2}\bar{\varphi}^{2}=-\frac{2\bar{N}_{\text{Tr}}\bar{g}^{2}r}{3\bar{\lambda}}.
\end{align}
To compute the mass gap ratio, we will use $m_{\psi,b}^{2}$ which is the heaviest one. 
Then, the mass gap ratio is given by
\begin{align}
\mathcal{R}_{G}={}&\frac{\bar{N}_{\text{Tr}}}{N}\frac{m_{\varphi}^{2}}{m_{\psi}^{2}}=\frac{2\lambda}{\alpha_{g}}.\label{eq:mass_gap}
\end{align}
For $N=2$, at the fixed point,
\begin{widetext}
\begin{align}
\mathcal{R}_{G}={}&\frac{(F_{2}(N_{\text{f}})-N_{\text{f}}+1)}{5}+\frac{((76 N_{\text{f}}^{3}+572 N_{\text{f}}^{2}-6467N_{\text{f}}+219)-(76N_{\text{f}}^{2}-872N_{\text{f}}-219)F_{2}(N_{\text{f}}))}{1000(N_{\text{f}}+1)F_{2}(N_{\text{f}})}\epsilon\\
&+\frac{\epsilon^{2}}{200000(N_{\text{f}}+1)^{3}F_{2}(N_{\text{f}})}\left[(304N_{\text{f}}^{5}-8960N_{\text{f}}^{4}-120600N_{\text{f}}^{3}+244420N_{\text{f}}^{2}-1284845 N_{\text{f}}+21681)\right.\notag\\
&\left.-(304N_{\text{f}}^{4}-14736N_{\text{f}}^{3}+214104N_{\text{f}}^{2}-315716N_{\text{f}}-21681)F_{2}(N_{\text{f}})\right]+\mathcal{O}(\epsilon^{3})
\end{align}
and for $N=3$,
\begin{align}
\mathcal{R}_{G}={}&\frac{3(F_{3}(N_{\text{f}})-N_{\text{f}}+2)}{13}
+\frac{((1884N_{\text{f}}^{3}+25174N_{\text{f}}^{2}-244555N_{\text{f}}+13710)-(1884N_{\text{f}}^{2}-20042N{f}-6855)F_{3}(N_{\text{f}}))}{17576(N_{\text{f}}+2)F_{3}(N_{\text{f}})}\epsilon\\
&+\frac{\epsilon^{2}}{23762752(N_{\text{f}}+2)^{3}F_{3}(N_{\text{f}})}\left[(52752 N_{\text{f}}^{5}-3767744 N_{\text{f}}^{4}-69585492 N_{\text{f}}^{3}+420724100 N_{\text{f}}^{2}-221946443 N_{\text{f}}-6841290)\right.\notag\\
&\left.-(52752N_{\text{f}}^{4}-5033792N_{\text{f}}^{3}+66312588N_{\text{f}}^{2}-64565476N_{\text{f}}+3420645)F_{3}(N_{\text{f}})\right]+\mathcal{O}(\epsilon^{3})
\end{align}
\end{widetext}

\begin{figure}
\includegraphics[width=0.9\linewidth]{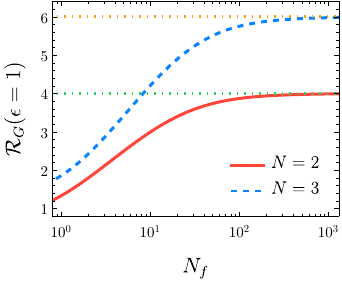}
\caption{
Mass gap ratio between boson and fermion masses with $N=2$ and 3 when $\epsilon=1$.
The red solid and blue dashed lines are $N=2$ and 3, respectively, and the green and orange dotted lines are 4 and 6.
$N=2$ approaches to 4 but $N=3$ goes to 6.
}\label{fig:massgap}
\end{figure}

For example, when $N_{\text{f}}=1$,
\begin{align}
\mathcal{R}_{G}={}
\begin{cases}
1.265+0.0648\epsilon-0.0274\epsilon^{2},&N=2\\
1.911-0.0565\epsilon+0.0251\epsilon^{2},&N=3\\
\end{cases}
\end{align}
and putting $\epsilon=1$,
\begin{align}
\mathcal{R}_{G}=
\begin{cases}
1.302,&N=2\\
1.874,&N=3.\\
\end{cases}
\end{align}
The result in terms of $N_{\text{f}}$ with $\epsilon=1$ is presented in Fig.~\ref{fig:massgap}.
For large $N_{\text{f}}$, $\mathcal{R}_{G}(N=2,N_{\text{f}}\rightarrow\infty)\rightarrow4$ and $\mathcal{R}_{G}(N=3,N_{\text{f}}\rightarrow\infty)\rightarrow6$, respectively. Hence, the case $N=2$ shows the same behavior in the large $N_f$ limit as in Ref~\cite{han2024b}, whereas the $N=3$ case is different. 
Note that the mass gap ratio with $m_{\psi,a}^{2}$ for $N=3$, instead of $m_{\psi,b}^{2}$ is given by $4\mathcal{R}_{G}(N=3)$ and it will approach 24 when $N_{\text{f}}\rightarrow\infty$ because $4m_{\psi,a}^{2}=m_{\psi,b}^{2}$.

\section{Conclusion and Discussion}\label{sec:conclusion}

We studied the GNY field theory with SO(2) and SO(3) real symmetric tensor order parameters coupled to Dirac fermions.
Up to the quartic terms in the bosonic sector of the action, the two symmetric tensor order parameters in question can alternatively be described by SO(2) and SO(5) vector order parameters, respectively. We computed the two-loop RG flow equations and found the stable (critical) interacting fixed point for any number of fermion flavors.
The values of the anomalous dimensions of fermions and order parameters and the correlation length exponent have been computed up to the order of $\epsilon^{2}$, as well as the mass gap ratio between fermions and order parameters in the ordered phase.

We have also computed the RG flow equations with two independent sextic self-interactions of the order parameter, which are the leading terms that reduce the accidental SO(5) symmetry back to SO(3), to determine the precise form of the broken symmetry configuration in this case. The result is the uniaxial nematic state, which reduces SO(3) to SO(2) and leads to a full gap in the fermion spectrum. This agrees with mean-field theory in the Gross-Neveu formulation \cite{ray2024}.

The examples of  $N=2$ and $N=3$ considered here differ from the GNY theories for the general case of $\text{SO}(N\geq 4)$ real symmetric tensor field theories in that, unlike there, the critical fixed points exist for any number of fermion flavors. This is a consequence of the fact that the theory admits only one quartic self-interaction term, and on that level of approximation resembles the GNY theories for SO(2) and SO(5) vector order parameters. Without fermions, the critical behavior of the two tensor theories is indeed identical to those of their vector counterparts. 
For $N=2$, the Yukawa theory is equivalent to the chiral XY model, and our results reduce to that universality class. For $N=3$, however, the critical behavior of the tensor Yukawa theory becomes different from the vector theory, and lies in a genuinely new universality class. This hence represents one, and to the best of our knowledge only, example of distinctly tensorial quantum criticality which exists for all numbers of fermion flavors. We hope that the predictions at $\mathcal{O}(\epsilon^2)$ for the critical exponents and mass gap ratio we have presented here for this universality class will be compared in the future with numerical simulations, conformal bootstrap \cite{Reehorst2023}, and actual experiments. Particularly on the simulations front, Ref.~\cite{Liu2024} recently presented estimates for critical exponents of the vectorial version of the Gross-Neveu-Yukawa SO(3) universality class and suggests that the tensorial universality class can also be accessed without encountering sign problems.

Our findings also have direct applications to other systems. For example, the results apply to $N_{\text{f}}$ copies of four-component Majorana fermions in $D=4$. By replacing $N_{\text{f}}\rightarrow N_{\text{f}}/2$, it corresponds to $N_{\text{f}}$ copies of two-component SO($N=2,3$) Majorana fermions in $D=3$ \cite{han2024b}. Conversely, by replacing $N_{\text{f}}\rightarrow 2N_{\text{f}}$, they apply to $N_{\text{f}}$ copies of four-component SO($N=2,3$) Dirac fermions in $D=4$. In particular, the two-component Majorana version of the model in $D=3$ has been proposed to be realized at the quantum phase transition of a SO(3) spin-orbital liquid to a quadrupolar-ordered phase \cite{ray2024}. Our predictions for the critical exponents of this transition are hence also of relevance to the study of fractionalized phases of matter in frustrated magnets.

\begin{acknowledgments}
The authors thank H.~Osborn and L.~Janssen for comments on the manuscript.
This work was supported by the NSERC of Canada. S.R.~acknowledges support from VILLUM FONDEN (grant no.~29405) and the Deutsche Forschungsgemeinschaft (DFG) through the Walter Benjamin program (Grant No.~RA3854/1-1, Project
ID No.~518075237) for parts of this project.
\end{acknowledgments}

%
\end{document}